\documentclass{iau} 
\usepackage{graphicx}

\title[JD 11.~~High Mass X-Ray Binary Microquasars in the Early Universe] 
{Black Hole High Mass X-ray Binary Microquasars at Cosmic Dawn}

\author[I.F. Mirabel]   
{I.F. Mirabel$^{1,2}$}

\affiliation{$^1$Institute of Astronomy and Space Physics. CONICET - Universidad de Buenos Aires,   
Ciudad Universitaria, Av. Cantilo S/N , 1428 Buenos Aires - Argentina. email: {\tt mirabel@iafe.uba.ar} 
\\[\affilskip]$^2$Laboratoire AIM-Paris-Saclay, CEA/DSM/Irfu/DAP−CNRS, CEA-Saclay, 
pt courrier 131, 91191 Gif-sur-Yvette, France. email: {\tt felix.mirabel@cea.fr}}

\pubyear{2019}
\volume{xxx}  
\jname{High-mass X-ray binaries: illuminating the passage from massive binaries to merging compact objects}
\editors{T. Bulik, E. Bozzo, D. Gies \& L. Oskinova, eds.}
\begin{document}

\maketitle

\begin{abstract}Theoretical models and observations suggest that primordial Stellar Black Holes (Pop-III-BHs) were prolifically formed in HMXBs, which are powerful relativistic jet sources of synchrotron radiation called Microquasars (MQs).

Large populations of BH-HMXB-MQs at cosmic dawn produce a smooth synchrotron cosmic radio background (CRB) that could account for the excess amplitude of atomic hydrogen absorption at z$\sim$17, recently reported by EDGES.

BH-HMXB-MQs at cosmic dawn precede supernovae, neutron stars and dust. BH-HMXB-MQs promptly inject into the IGM hard X-rays and relativistic jets, which overtake the slowly expanding HII regions ionized by progenitor Pop-III stars, heating and partially ionizing the IGM over larger distance scales.

BH-HMXBs are channels for the formation of Binary-Black-Holes (BBHs). The large masses of BBHs detected by gravitational waves, relative to the masses of BHs detected by X-rays, and the high rates of BBH-mergers, are consistent with high formation rates of BH-HMXBs and BBHs in the early universe.

\end{abstract}

\keywords{X-rays: binaries, microquasars, early universe, (cosmology:) black hole physics, gravitational waves}

\firstsection 
\section{HMXBs in the heating and reionization epochs: X-rays}

It is well established that between 380.000 and 1 billion years after the Big Bang the IGM underwent
a  phase transformation from cold and fully neutral to warm ($\sim$10$^{4}$ K) and ionized.
 Whether this phase transformation was fully driven and completed by photoionization from young hot stars 
is a question of topical interest in cosmology. \cite[Mirabel et al. (2011)]{Mirabel_etal11} and Fragos et al. (2013) proposed that besides the UV radiation from massive stars of populations III and II, and the soft X-rays from core-collapse SNe (Furlanetto et al. 2004), 
BH-HMXBs, the remnants of the first generations of massive binary stellar systems, likely played an important role in the process of heating, and possibly a secondary, complementary role to the reionization 
of the IGM that took place $\leq$10$^9$ years after the Big Bang. Because X-ray photons from HMXBs have longer mean free paths than UV photons from their massive stellar progenitors, 
the X-rays and possibly the relativistic jets from BH-HMXB-MQs formed from Pop III  and Pop II stars would have heated the IGM across larger volumes of space during reionization. In fact, X-ray photons  
ionize hydrogen and helium, and the free particles deposit their kinetic energy in the IGM by secondary ionization and free-free heating.

An extensive presentation of the theoretical and observational grounds for the hypothesis of large populations of BH-HMXRBs in the early universe was published by \cite[Mirabel et al. (2011)]{Mirabel_etal11}. From that study was concluded the following:

1. The ratio of BHs to neutron stars (NS) and the ratio of BBHs  to solitary BHs should 
increase with redshift; that is, the rate of formation of BH-HMXBs is significantly
larger in the early Universe than at present.

2. Feedback from a Galactic BH-HMXBs during a typical whole
lifetime of 10$^7$ years is $\sim$ 3 $\times$10$^{52}$ erg, $\sim$30 times larger than the photonic and baryonic energy from a typical core collapse supernova. However, 
due to the low metallicities of the stellar progenitors BH-HMXBs in the early universe are likely more energetic. 

3. An accreting BH in a HMXB emits a total
number of ionizing photons that is comparable to its progenitor
star, but one X-ray photon emitted by an accreting BH  may cause the ionization of several tens of hydrogen
atoms in a fully neutral medium.

4. The most important effect of BH-HMXBs in the early universe
could be heating of the IGM. Soft X-rays and inverse-Compton scattering from relativistic electrons produced by
BH-HMXBs could heat the low-density medium over large volumes
to temperatures of $\sim$10$^4$ K, which would limit the recombination
rate of hydrogen keeping the IGM ionized.

5. A temperature of the IGM of $\sim$10$^4$ K would limit the formation of
faint galaxies at high redshifts. It constrains the total mass of
dwarf galaxies to $\geq$10$^9$ M$_{\odot}$, leaving 
dark matter halos of $\leq$10$^9$ M$_{\odot}$ with no barionic mass.

6. Therefore, BH-HMXBs in the early universe could be important ingredients
for reconciling the apparent disparity between the observed
number of faint dwarf galaxies in the Galactic halo with the much larger 
number of low-mass galaxies predicted by the cold dark matter
model of the universe.

7. An additional effect of metallicity in the formation of
BH-HMXBs  is to boost the formation of BBHs (\cite[Mirabel 2010]{Mirabel_10}), 
as the more likely first detected sources of gravitational waves
than NS-NS systems (\cite[Belczynski et al. 2010] {Belczynski_etal10}). 
These predictions have been confirmed by the first observations of the LIGO-Virgo collaboration.

\vskip .1in

After the publication by Mirabel et al. (2011) the observational grounds for the hypothesis of a high rate of HMXB formation in the early universe has been re-enforced by new observational results (e.g. \cite[Fragos et al. (2013)]{Fragos_etal13}; \cite[Basu-Zych et al. (2013)]{Basu-Zych_etal13}; \cite[Kaaret (2014)]{Kaaret14}; \cite[Douna et al. (2015)]{Douna_etal15}; \cite[Lehmer et al. (2016)]{Lehmer_etal16}; \cite[Brorby et al. (2016)]{Brorby_etal16}). \cite[Lehmer et al. (2016)]{Lehmer_etal16} found an increase in LMXB and HMXB
scaling relations with redshift as being due to declining host galaxy stellar ages and metallicities, respectively, and discussed how emission from HMXRBs could provide an important source of heating to the IGM in the
early universe, exceeding that of active galactic nuclei (AGN), as shown in Figure 1.

\begin{figure}[h]
\begin{center}
 \includegraphics[width=5.3in]{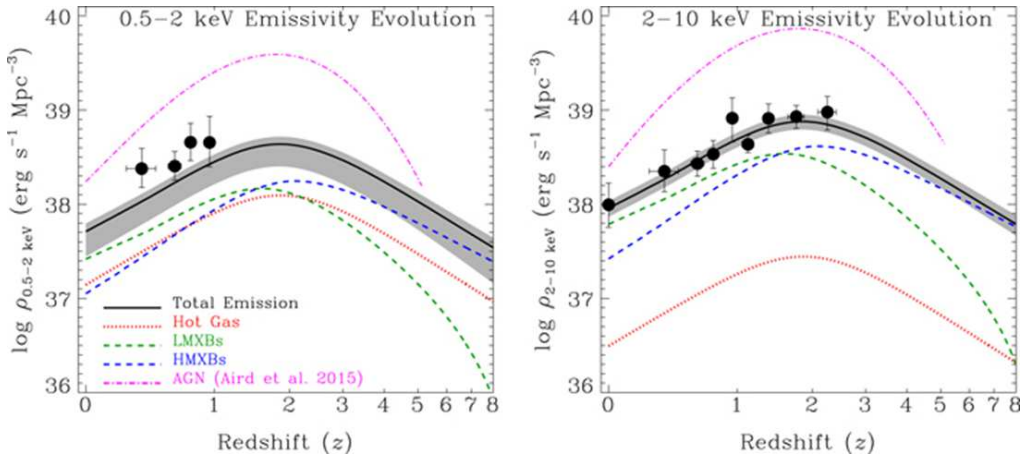} 
 \caption{\textbf: These figures from \cite[Lehmer et al. (2016)]{Lehmer_etal16} show how the volume-averaged emission of X-rays from different X-ray sources evolves over cosmic
time. The AGN curve, shown in magenta, decreases abruptly at z$\geq$6, while the total emission from galaxies (shown by the black solid line and gray shaded region)
dominates at higher z. The high mass X-ray binaries (blue
dashed line) are the main contributors of X-rays within galaxies at z$\geq$6.}

   \label{fig1}
\end{center}
\end{figure}

\section{X-rays, Gamma-rays, and relativistic jets from HMXB-MQs}

Most previous works on the impact of HMXBs in the heating and reionization epochs of the universe have considered feedback in the form of X-ray radiation. However, in recent years we have come to the realization that HMXBs are also Microquasar sources of relativistic jets and massive outflows  that dissipate a large fraction of the liberated accretion power in the form of relativistic outflows (\cite[Mirabel \& Rodr\'\i guez 1999)]{Mirabel \& Rodr\'\i guez98}. In fact, the best studied HMXBs in our Galaxy, Cygnus X-1, SS433 and Cygnus X-3, are sources of powerful jets that are equally energetic or even more energetic, than their photonic feedbacks (\cite[Mirabel \& Rodr\'\i guez 1999)]{Mirabel \& Rodr\'\i guez99}. High energy gamma-ray emissions have also been reported from these three HMXB-MQs, and other HMXB classes (short reviews in Mirabel 2006 \& Mirabel 2012) 

Associated with Cygnus X-1, Gallo et al. (2005) found at a distance of $\sim$5 pc from the BH-HMXB a ring-like shock structure produced by the jets from the accreting BH, with a kinetic energy injection of at least that of the total X-ray luminosity. GeV emission from Cygnus X-1 most likely associated with the relativistic jets was reported by Zanin et al. (2016). 

In SS433 relativistic jets moving at $\sim$0.26c carry heavy nuclei with a mechanical energy $\geq$10$^{39}$erg s$^{-1}$. They impact at a distance of several tens of parsecs on the interstellar medium and inflate laterally the W50 nebula that hosts the HMXB (Mirabel \& Rodr\'\i guez 1999). Recently, The High Altitude Water Cherenkov (HAWC 2018) reported from SS433/W50 detection of gamma-ray emission of at least 25 TeVs, from the lobes of W50
in which the jets terminate about 40 pc from the central source. 

Cygnus X-3 is at a larger distance of ‎7.4$\pm$1.1 kpc, the donor star shows Wolf-Rayet features and the orbital period is 4.8 hs. This HMXB-MQ is one of the most powerful X-ray transient sources of relativistic radio jets in the Milky Way, which produce synchrotron fluxes at radio waves of up to several tens of Jy. This source has been detected by the gamma-ray satellites AGILE (Tavani et al. 2009) and Fermi-LAT (2009). Cygnus X-3 is in a crowded region and the later secured that the high energy emission actually comes from Cygnus X-3, by the orbital period in gamma rays, as well as with the correlation of the LAT flux with radio emission from the relativistic jets.  

On the other hand, in nearby dwarf galaxies are found inflated cavities by relativistic jets and massive outflows from HMXBs (Feng \& Soria 2011). For instance, Pakull, Soria \& Motch (2010) found in NGC 7793 a jet-inflated bubble, with a diameter of 300 pc, surrounding a HMXB-MQ. Pakull et al. (2010) estimated for this HMXB a mechanical kinetic energy injection of $\sim$10$^{4}$ that of X-rays. 

\vskip .05in

\cite[Heinz \& Sunyaev (2002)]{Heinz \& Sunyaev02} studied the physics of the interaction of jets from MQs with the interstellar medium (ISM) showing that cosmic rays are produced by shocks of the relativistic jets with the ISM. The relativistic particles of the jets deposit a fraction of their kinetic energy at the interface between the jet and the ambient medium (working surface), into random, isotropic particle energy, likely contributing to the heating of the IGM.  Following that work \cite[Tueros et al. (2014)]{Tueros_etal14} proposed that microquasar jets contribute to the reionization of the IGM at cosmic dawn. More recently, \cite[Douna et al. (2018)]{Douna_etal18} by Monte Carlo simulations find that the contribution of microquasar jets to the heating of the IGM  is of the same order of magnitude as that of cosmic rays from SNe.

\vskip .1in
\section{Stellar black holes formed by implosion}

Theoretical models predict that massive stars with decreasing metallicity produce weaker stellar winds, allowing larger numbers of massive stellar binaries to become HMXBs, with more massive compact objects formed by direct collapse (BHs), and with more luminous donor stars.  Therefore, it is expected that HMXBs formed in the low metallicity environments of the early universe should be more numerous and more energetic than HMXBs in the local universe. 

The question on how stellar black holes are formed is of topical interest for the incipient Gravitational-Wave Astrophysics. Whether BHs are formed through energetic natal supernova kicks or by implosion may impact the final evolutionary stage of a large fraction of massive binary stars, the numbers of BBHs that will be formed, and therefore the merger rates of BBHs that will be detected in GWs observations by the LIGO-Virgo collaboration and other GW research collaborations. In section 7 is discussed the formation of BHs by implosion in the context of the different channels for the formation of BBHs.

From population synthesis models it is inferred that the disruption rate of non-tight massive stellar binaries increases by two orders of magnitude varying from the assumption of BH formation with no kicks to that of a kick distribution typical of neutron stars (\cite[Dominik et al. 2012]{Dominik_etal12}). Besides, the escape velocity from a globular cluster of $\leq$10$^7$ M$_{\odot}$ is few tens km s$^{-1}$, and  if formed with a kick distribution typical of neutron stars most BHs would be kick out from globular clusters by SN natal kicks. 

Theoretical models set progenitor masses for BH formation by implosion, but observational evidences have been elusive.  The kinematics of BH-X-ray binaries can provide observations to contrast the theories on the formation of BHs. If a compact object is accompanied by a mass-donor star in an X-ray binary, it is possible to determine the distance, proper motion, and radial velocity of the center of mass of the system, from which can be derived the velocity in three dimensions of space, and in some cases, also infer the site of birth of the BH. Here are summarized the recently improved determination, in most cases by VLBI at radio wavelengths, of the kinematics of two Galactic BH-X-ray binaries formed by direct collapse. When available, short comments are given on preliminary results obtained with GAIA.

\textbf{Cygnus X-1} is a X-ray binary at a distance of 1.86 $\pm$ 0.1 kpc  composed by a BH of 14.8 $\pm$ 1.0 M$_{\odot}$  and a 09.7lab donor star of 19.2 $\pm$1.9 M$_{\odot}$ with an orbital period of 5.6 days and eccentricity of 0.018 $\pm$ 0.003.  Cygnus X-1 appears to be at comparable distance and moving together with the association of massive stars Cygnus OB3 (\cite[Mirabel \& Rodrigues 2003)]{MirabelRodriges 2003}. Therefore, it had been proposed that the BH in Cygnus X-1 was formed in situ and did not receive an energetic trigger from a natal or nearby supernova. 

More recently, a trigonometric parallax that corresponds to a distance of 1.86$\pm$0.12 kpc was obtained with the Very Long Baseline Array by Reid et al. (2011). These authors also measured the proper motion of Cygnus X-1 which, when coupled to the distance and Doppler shift, gives the three-dimensional space motion of the system. Reid et al. (2011) conclude that the binary did not experience a large "kick" at BH formation. 

\begin{figure}[h]
\begin{center}
 \includegraphics[width=5.3in]{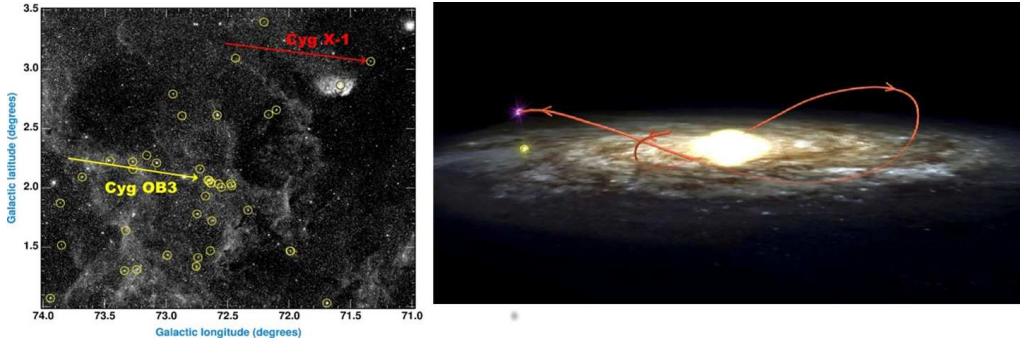} 
 \caption{\textbf{Left}: Optical image of the sky around the BH X-ray binary Cygnus X-1 and the association of massive stars Cygnus OB3. The red and yellow arrows show the magnitudes and directions of the motion in the plane of the sky of the radio counterpart of Cygnus X-1 and the average Hipparcos motion of the massive stars of Cygnus OB3 (circled in yellow) for the past 0.5 Millions years, respectively. After the formation of the BH, Cygnus X-1 remained anchored in the parent association of massive stars Cygnus OB3.       From \cite[Mirabel \& Rodrigues (2003)]{MirabelRodriges 2003}. \textbf{Right}: Schematic Galactic orbit of the Galactic Halo BH X-ray binary XTE J1118+480 (red curve) during the last $\sim$240 Myr, which corresponds to the orbital period of the Sun around the Galactic Center. The source left the plane towards the Northern Galactic hemisphere with a galactic-centric velocity of 348$\pm$18 km s$^{-1}$, which after subtraction of the velocity vector due to Galactic rotation, corresponds to a peculiar space velocity of 217$\pm$18 km s$^{-1}$  relative to the Galactic disk frame, and a component perpendicular to the plane of 126$\pm$18 km s$^{-1}$. The galactic orbit of XTE J1118+480 has an eccentricity of 0.54. At the present epoch XTE J1118+480 is at a distance from the Sun of only 1.9$\pm$0.4 kpc flying through the Galactic local neighborhood with a velocity of 145 km s$^{-1}$. From Mirabel et al. (2001)}

   \label{fig2}
\end{center}
\end{figure}

On the other hand, the comparison with the results from a new reduction of the Hipparcos data by Mel’nik and Dambis (2009) to infer the mean distance and proper motion of Cygnus OB3, reaffirmed the conjecture that Cygnus OB3 is the parent association of Cygnus X-1. From a preliminary analysis of the GAIA second data release it is found that in fact Cygnus OB3 most likely is the parent association of Cygnus X-1 (\cite[Garc\'\i a et al. 2019]{Garc\'\i a_etal19}) as proposed by \cite[Mirabel \& Rodrigues (2003)]{MirabelRodrigues 2003}.  However, so far there is a discrepancy between the GAIA and the VLBI radio parallaxes of Cygnux X-1, that could be attributable either to intrinsic orbital wobble, or to systematic pipeline measurement uncertainties in the GAIA data (\cite[Gandhi et al. 2019] {Gandhi_etal19}). The motions on the plane of the sky of Cygnus X-1 and Cygnus OB3 are shown on the left side of  Figure \ref{fig2}. 

The upper limit of the velocity in three dimensions of Cygnus X-1 relative to the mean velocity of Cygnus OB3 is 9 $\pm$ 2 km s$^{-1}$, which is typical of random velocities of stars in expanding associations of massive stars. 

From the equations for spherical mass ejection at BH formation in massive stellar binaries, it is estimated that the maximum mass that could have been suddenly ejected to accelerate the binary without binary disruption to a velocity of 9 $\pm$ 2 km s$^{-1}$ is $\leq$1$\pm$0.3 M$_{\odot}$. Indeed, there are no observational evidences for a SN remnant in the radio continuum, X-rays, and atomic hydrogen surveys of the region where Cygnus X-1 was most likely formed. 

Mirabel \& Rodrigues (2003) estimated that the initial mass of the progenitor of the BH is $\sim$40 $\pm$ 5 M$_{\odot}$ which may have lost $\sim$25 M$_{\odot}$ by stellar winds during a Wolf-Rayet stage and mass exchange with the presently massive stellar donor. The observational mass lower limit of $\sim$40 M$_{\odot}$ for the progenitor is the same as the mass theoretically predicted for the transition from fall-back to complete collapse for a BH progenitor of solar metallicity (\cite[Fryer et al. 2012]{Fryer_etal12}), and the complete collapse of stellar helium cores in stars of solar metallicity (\cite[Sukhbold et al. 2016]{Sukhbold_etal16}).  

\textbf{GRS 1915+105} is a low-mass X-ray binary containing a BH of 10.1 $\pm$  0.6 M$_{\odot}$ and a donor star of spectral type K-M III of 0.5 $\pm$  0.3 M$_{\odot}$, with a 34 day circular orbital period. The companion overflows its Roche lobe and the system exhibits episodic superluminal radio jets (Mirabel \& Rodr\'\i guez 1994). 

From a decade of astrometry with the NRAO Very Long Baseline Array it was determined a parallax distance of of 8.6 $\pm$ 1.8 kpc and proper motion, that together with the published radial velocity, is determined a modest peculiar velocity of 22 $\pm$  24 km s$^{-1}$  (\cite[Reid et al. 2014]{Reid_etal14}), which is consistent with the earlier proposition that the BH in GRS 1915+105 was formed without a strong natal kick, like the 14.8 $\pm$ 1.0 M$_{\odot}$ BH in Cygnus X-1 (\cite[Mirabel et al. 2001]{Mirabel_etal01}). 
The modest peculiar speed of 22 $\pm$ 24 km s$^{-1}$  at the parallax distance and a donor star in the giant branch suggest that GRS 1915+105 is an old system that has orbited the Galaxy many times, acquiring a peculiar velocity component on the galactic disk of 20-30 km s$^{-1}$, consistent with the velocity dispersions of $\sim$20 km s$^{-1}$ of old stellar systems in the thin disk, due to galactic diffusion by random gravitational perturbations from encounters with spiral arms and giant molecular clouds. 

\vskip.1in

In summary, the kinematics in three dimensions of Cygnus X-1 relative to the parent association of massive stars Cygnus OB3, and the kinematics of GRS 1915+105 relative to its Galactic environment, suggest, irrespective of their origin in isolated massive stellar binaries or in typical dense clusters of less than 10$^7$ M$_{\odot}$, that the black holes in the X-ray binaries Cygnus X-1 and in GRS 1915+105, were formed in situ by complete or almost complete collapse of massive stars, with no energetic kicks from natal supernovae. These observational results are consistent with theoretical models, and in particular, with one of the most recent models, where the black holes of $\sim$15 M$_{\odot}$  in Cygnus X-1, and $\sim$10 M$_{\odot}$ in GRS 1915+105, were formed respectively by complete, and almost complete collapse of stellar helium cores (\cite[Sukhbold et al. 2016]{Sukhbold_etal16}). 

\section{Black hole X-ray binaries with peculiar velocities}

\textbf{GRO J1655-40} is a X-ray binary with a BH of 5.3 $\pm$  0.7 M$_{\odot}$ and a F6-F7 IV donor star.  It had been estimated that this BH-XRB is moving with a peculiar velocity with respect to its environment of 112 $\pm$ 18 km s$^{-1}$, in a highly eccentric (e=0.34 $\pm$ 0.05) Galactic orbit (\cite[Mirabel et al. 2002]{Mirabel_etal02}). 

From a preliminary analysis of the GAIA second data release it is estimated that the parallax-inversion distance of this X-ray binary corresponds to a distance of 3.66$\pm$1.01 kpc, and the peculiar velocity relative to Galactic rotation would be 150.6$\pm$13 km s$^{-1}$  (\cite[Gandhi et al. 2019] {Gandhi_etal19}). The runaway linear momentum and kinetic energy of this X-ray binary are similar to those of solitary runaway neutron stars and millisecond pulsars with the most extreme runaway velocities. 

\textbf{XTE J1118+480} is a high-galactic-latitude (l=157$^{o}$.78, b=+62$^{o}$.38) X-ray binary with a BH of 7.6 $\pm$ 0.7 M$_{\odot}$ and a 0.18 M$_{\odot}$ donor star of spectral type K7V-M1V. It is  moving in a highly eccentric orbit around the Galactic center region, as some ancient stars and globular clusters in the halo of the Galaxy. On the right two panels of Figure 2 are shown the top and side views of the orbital path of this X-ray binary relative to the Galactic disk (\cite[Mirabel et al. 2001]{Mirabel_etal01}). 

\textbf{V404 Cyg (GS 2023+338)} is a low mass X-ray binary system composed of a BH of 9.0 $\pm$ 0.6 M$_{\odot}$  and a 0.75 M$_{\odot}$ donor of spectral type K0 IV. From astrometric VLBI observations, it was measured for this system a parallax distance of 2.39 $\pm$ 0.14 kpc (\cite[Miller-Jones et al. 2009a]{Miller-Jones_etal09a}). Together with the proper motion Miller-Jones et al. (2009b) derived a peculiar velocity of 39.9 $\pm$ 5.5 km s$^{-1}$, with a component on the Galactic plane of 39.6 km s$^{-1}$, that is $\sim$2 times larger than the expected velocity dispersion in the Galactic plane. 

From a preliminary analysis of the GAIA second data release Gandhi et al. (2018) estimated a parallax-inversion distance of 2.28$\pm$0.52 kpc and a peculiar velocity relative to Galactic rotation of 51.5$\pm$16.0 km s$^{-1}$, which -within the errors- are consistent with the radio VLBI observations.

\vskip.1in

\textbf{On the interpretation of the anomalous velocities of black hole X-ray binaries}.
GRO J1655-40 and V404 Cyg are in the Galactic disk (b $\leq$ 3.2$^{o}$; distances from the plane z $\leq$ 0.15 kpc), where were likely formed. XTE J118+480 is in the Galactic halo (b = 62.3$^{o}$;  z = 1.5 kpc) and could either have been propelled to its present position from the Galactic disk by a supernova explosion, or have been formed in a globular cluster from which it could have escaped with a mild velocity of few tens of km s$^{-1}$. 

If these X-ray binaries were formed from binary stars in a field of relative low density, one would conclude that the three BHs with $\leq$10 M$_{\odot}$ were formed with significant natal kicks, whereas the BHs with $\geq$10 M$_{\odot}$ in Cygnus X-1 and GRS 1915+105 were formed with no energetic natal supernovae kicks. 

However, if these "runaway" BH-XRBs were formed in dense stellar clusters, the anomalous velocities of the XRBs barycenter could have been caused by dynamical interactions in the stellar cluster, or by the explosion of a nearby massive star, and not necessarily by the explosion of the progenitor of the "runaway" BH. Furthermore, the possible supernova nuclear-synthetic products in the atmospheres of the donor stars in GRO J1655-40, XTE 1118+480 and V404 Cyg, could be due to contamination by the explosion of a nearby star in a high density natal environment, rather than to the explosion of the BH progenitor. 

In fact, the three BH "runaway" XRBs have low mass donors, and their linear momenta are comparable to those of runaway massive stars likely ejected from multiple stellar systems by the Blaauw mechanism. Therefore, without knowing the origin of a "runaway" X-ray binary, it is not possible to constrain from its peculiar velocity alone, the strength of a putative natal supernova kick to the compact object in the XRB. It is expected that future parallax distances and proper motions of BHs and their environment, determined at radio wavelengths by VLBI and at optical wavelengths with GAIA, will provide further observational constraints on the physical mechanisms of stellar BH formation.

\section{Signatures of BH-HMXB-MQs in the earlier universe}

{\bf Stellar BHs in the Local Universe.} 

It is estimated that in the Galaxy there have been formed 10$^{8}$ to 10$^{9}$ stellar-mass BHs (e.g. \cite[Brown \& Bethe 1994]{Brown \& Bethe94}; \cite[Timmes et al. 1996]{Timmes_etal96}). Assuming a mass of 10 M$_{\odot}$ for each BH, the integrated mass of those stellar BHs would be $\sim$10$^3$ times the mass of the supermassive BH of $\sim$4$\times$10$^6$ M$_{\odot}$ at the center of the Galaxy. The majority of stellar BHs of that large population presently are inactive, but in earlier epochs of Galactic evolution a significant fraction may have had periods of strong feedback activity. Until now out of that putative large population of stellar BHs only a few tens of BH-XRBs have been dynamically identified, Cygnus X-1, and likely Cygnus X-3 and SS 433, being the best studied Galactic HMXB-MQs believed to harbor stellar BHs. 

BH-HMXBs are sources of powerful collimated relativistic jets and massive outflows, capable of blowing large two sided cavities in the ISM (Mirabel \& Rodr\'\i guez 1999; Gallo et al. 2005). Similar HMXB-MQs have been identified in nearby dwarf galaxies of relative low metallicity (\cite[Feng \& Soria 2011]{Feng \&  Soria11}). As shown in sections 1 and 2, it is expected that a large fraction of the immediate remnants of the first massive stars in the early universe are BH-HMXB-MQs, with powerful relativistic jets that generate synchrotron radiation (\cite[Mirabel \& Rodr\'\i guez 1999]{Mirabel \& Rodr\'\i guez99}). In this context one may ask whether that large population of BH-HMXB-MQs at cosmic dawn could have produced a diffuse cosmic radio synchrotron background, and whether it could have already been observed.

\vskip 0.05in

{\bf A smooth radio synchrotron background of unknown origin?} 

From observations with a balloon-borne instrument called ARCADE 2, \cite[Fixsen et al. (2011)]{Fixsen_etal11} reported an excess radio synchrotron background, consistent with the Cosmic Microwave Background (CMB) radiation at high frequencies, but significantly deviating from a blackbody spectrum at low frequencies (see Figure 3-Left). A possible radio background had been suggested before by Bridle (1967), and its possible existence was later reaffirmed with ARCADE 2 and other observations (Singal et al. 2018 and references therein). 

The origin of this radio synchrotron background excess is a subject of debate because the observed level of surface brightness is substantially larger than expected from observed radio counts, unresolved emission from the known radio point source population, the spectrum of this excess emission being flatter than the much deeper spectrum of radio point sources, and it cannot be explained by CMB spectral distortions (Seiffert et al. 2011). 

Condon et al. (2012) used the Very Large Array (VLA) to image at 3 GHz a primary beam at high resolution and high sensitivity. They conclude from radio source count considerations that neither AGN-driven sources or star-forming galaxies can provide the bulk of the observed level of the radio synchrotron background reported by the observations with ARCADE 2. Condon et al. (2012) estimate that any new discrete-source population able to produce such
a bright and smooth background is far too numerous to be associated with galaxies brighter than m$_{AB}$=+29. 

It was believed that this reported background radiation may originate from very early cosmic times because it does not seem to be associated with far-infrared thermal emission from dust. If it were produced by sources that follow the correlation between radio emission and the far-infrared radiated by dust observed in galaxies, the cosmic far-infrared background (CIB) would be overproduced above the observed levels with Planck (Ysard \& Lagache 2012). Although the FIR-Radio correlation of galaxies may evolve with redshift (Ivison et al. 2010), Magnelli et al. (2015) conclude that star burst galaxies responsible for the CIB contribute $\leq10$\% of a putative cosmic radio background (CRB). In this context, $\geq$90\% of a CRB would come from sources at cosmic dawn that produce radio synchrotron emission, but do not produce radiation by dust. 

Caveats on the existence of that smooth radio background have been formulated by Vernstrom et al. (2015). However, Dowell and Taylor (2018) studied more recently the radio background between 40 and 80 MHz using the all-sky maps from the LWA1 Low Frequency Sky Survey (LLFSS). They confirm the strong, diffuse radio background suggested by the ARCADE 2 observations in the 3 to 10 GHz range. Dowell and Taylor (2018) modeled this radio background by a power law with a spectral index of −2.58$\pm$0.05 and a temperature at the rest frame 21 cm frequency of 603$\pm$97 mK. But Subrahmanyan \& Cowsik (2013)  claim that instead modeling Galactic emission as a plane parallel
slab, a more realistic modeling yields estimates for the smooth extragalactic brightness that are consistent with
expectations from known extragalactic radio source populations.

\begin{figure}[h]
\begin{center}
 \includegraphics[width=5.3in]{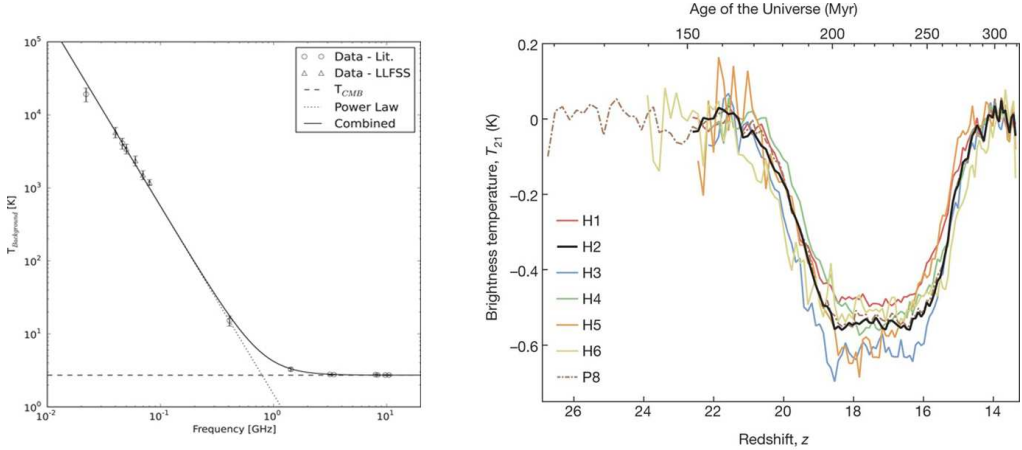} 
 \caption{\textbf{Left}: Modeled extragalactic temperature as a function of frequency from LLFSS and other maps by Dowell \& Taylor (2018). The solid line shows the best fit to the sum of a power law (dotted line) with a spectral index of −2.58 and the CMB (dashed line) at 2.722 K. 
\hspace{30pt} \textbf{Right}: Best-fitting 21-cm absorption profiles for each hardware case of EDGES by Bowman et al. (2018a).
Each profile for the brightness temperature is plotted against redshift z and the corresponding age of the Universe.
The thick black line is the model fit for the hardware and analysis
configuration with the highest signal-to-noise ratio, processed using 60 to 99 MHz and a four-term polynomial
for the foreground model.}

   \label{fig3}
\end{center}
\end{figure}

{\bf A deep HI absorption at z$\sim$17.} 

Figure 3-Right shows a deep absorption feature of $\sim$0.5 K centered at 78 MHz (z=17) in the averaged sky of the redshifted 21 cm spectrum of atomic hydrogen found by the Experiment to Detect the Global EoR Signatures (EDGES) from Bowman et al. (2018a). The absorption feature is $\geq$2 times stronger than the 21-cm signal absorption expected to be observed against the Cosmic Microwave Background (CMB) by standard astrophysical models of cosmic dawn (e.g. Figure 4-Right in next section). 

The search for redshifted 21 cm radiation of HI at cosmic dawn between 50 and 200 MHz require a careful and accurate subtraction of the sky foreground, since the searched signal is orders of magnitude below the foreground. Hills et al. (2018) expressed concerns about the modelling of the EDGES data in the reported absorption by Bowman et al. (2018a). These concerns were discussed in a reply by Bowman et al. (2018b). While waiting for an independent detection of this absorption signal, in the following it will be tentatively assumed that it is produced by neutral gas at cosmic dawn.  

Madau et al. (1997) studied the dependence of the HI signal amplitude $\delta T_{b}$ on the spin temperature T$_{s}$, the temperature of the Cosmic Microwave Background T$_{CMB}$, and an additional possible background T$_{rad}$. Following that study the amplitude of the HI absorption can be expressed as:

\begin{equation}
\delta T_b \propto \left(1 - \frac{T_{\rm{CMB}} + T_{\rm{rad}}}{T_{\rm{s}}}\right)
\end{equation}

One interpretation of this unexpected large and frequency wide absorption was the decreased of T$_{s}$ by an increased cooling of the HI gas by  interaction with dark matter (e.g. Barkana 2018). 

However, Feng \& Holder (2018) propose an alternative interpretation, based on the possible existence  of an additional cosmic radio background (CRB) field component from cosmic dawn, to that of the Cosmic Microwave Background (CMB). A CRB would increase the brightness temperature of the background radiation field (T$_{rad}$ in eq. 5.1) seen by the absorbing atomic hydrogen that enshrouds the sources of radio synchrotron emission.  In this context, a CRB (Figure 3-Left), and the enhanced deep and wide 21cm absorption reported by EDGES (Figure 3-Right) would be physically related. In section 7 it is proposed that the most likely predominant sources of a CRB are the BH-HMXB-MQs remnants of Pop III stars.     

\section{Tomography of HI in the early universe: theory vs. observations}

One of the main scientific motivations of the Square Kilometer Array (SKA) and several other radio astronomy projects
for low frequency observations is the spatial and timing evolution of the redshifted hyperfine transition of atomic hydrogen at 21cm. HI encodes the thermal and ionization state of the IGM due to UVs from massive stars, X-rays, and cosmic rays. From early researches it was inferred that if the reionization and thermal state of the IGM are fully driven and completed by photoionization from young hot stars, the tomography of HI should exhibit a morphology with prominently marked frontiers between the neutral and ionized regions, namely, the morphology should look like the morphology of Swiss cheese. 

The tomography of cosmic heating and reionization inferred from the evolution of the redshifted 21cm line of atomic hydrogen for different values of the IGM temperature at redshifts z$\geq$6 was modeled incorporating feedback in the X-rays from different types of sources by several groups of researchers. In particular, Mirabel et al (2011) proposed BH-HMXB-MQs as such sources, and  Haiman (2011) remarked that if hard X-rays from these sources are important, the tomography of HI should be smoother than previously thought. 

Figure 4 produced by Pritchard and published in Mirabel et al. (2011), shows the expected HI brightness temperature as a function of frequency and redshift for different values of f$_{X}$, a variable derived from Furlanetto (2006), which denotes the fraction of the total luminosity that emerges to the IGM in the assumed relatively hard 2-10 keV band radiation of BH-HMXB-MQs. 

\begin{figure}[h]
\begin{center}
 \includegraphics[width=5.3in]{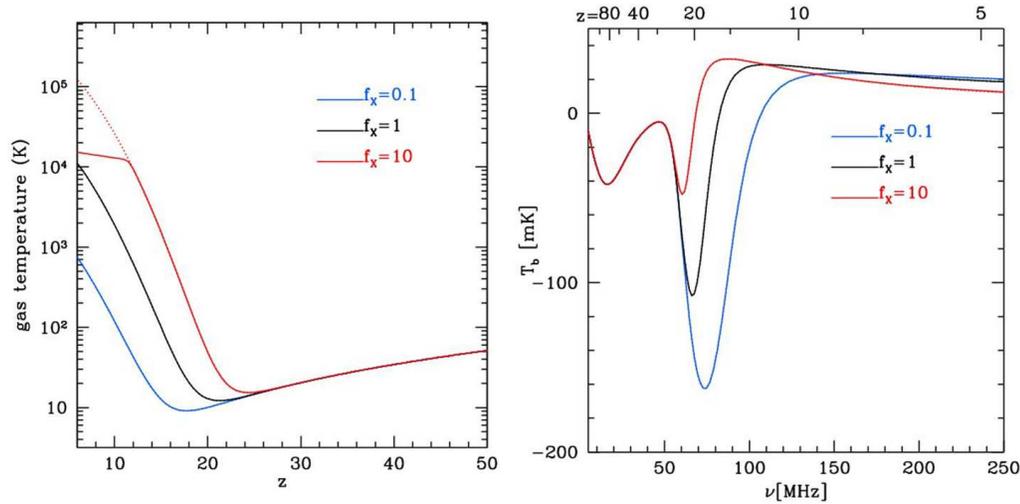} 
 \caption{\textbf{Left}: Thermal history of the neutral intergalactic medium (IGM) including heating by BH-HMXBs. The gas temperature of the IGM is plotted as a function of redshift z for three possible values of f$_{X}$ as defined by Eq. (6) in Mirabel et al. (2011). \textbf{Right}: Predicted brightness temperature of the hyperfine transition of the ground state of atomic hydrogen of 21 cm, averaged over the entire sky, as a function of redshift z, for the same three different values of f$_{X}$.  In this model the predicted frequency of the largest absorption amplitude against the Cosmic Microwave Background (CMB) for f$_{X}$$\sim$0.1 is at $\sim$73 MHz and the gas temperature $\sim$10 K. These values are close to the central frequency of 78 MHz (z$\sim$17) of the absorption reported by EDGES (Fig. 3-Right), which would correspond to f$_{X}$$\leq$0.1, IGM gas temperatures $\leq$10 K, and hydrogen densities N$_{H}$$\geq$5$\times$10$^{23}$ cm$^{-2}$. Both figures reproduced from Mirabel et al. (2011).}

   \label{fig4}
\end{center}
\end{figure}

A comparison of the model predictions by Mirabel et al. (2011) in Figure 4-Right, with the EDGES observed absorption in Figure 3-Right, shows that the best agreement of the model with observations would be for f$_{X}$$\leq$0.1, concerning the redshift of the maximum amplitude of the Gaussian-shape deep absorption and the overall frequency extension of the absorption. 

But there are striking differences in the amplitude and shape of the absorption profile between the expected Gaussian profile in Figure 4-Right, and the bottom-flat profile reported by EDGES in Figure 3-Right. The excess amplitude of the observed absorption may be due to the existence of a CRB. The non-Gaussian, flat bottom-shape of the EDGES signal can be accounted for by a normalized 21-cm bi-spectrum in fully-numerical simulations of the IGM heated by stellar sources and HMXBs (Watkinson et al. 2018). 

f$_{X}$$\leq$0.1 implies that the space density of the hydrogen that enshrouds the sources of X-rays and synchrotron radio emission should be high enough to block the X-rays, but not the radio synchrotron emission from the sources. Ewall-Wice et al. (2018) point out that in order to
avoid heating the IGM over the EDGES trough, the sources that produce the X-rays would need to be obscured by
hydrogen column depths of N$_{H}$$\sim$5$\times$10$^{23}$ cm$^{-2}$. In fact, N$_{H}$ column densities even larger than these are observed in star forming regions in the Milky Way, as well as in the central regions of extreme starburst far-infrared galaxies (Sanders \& Mirabel 1996). 

If the UV and X-ray photon escape fraction from stars and BH-HMXB-MQs of Pop III is f$_{esc}$$\leq$0.1, the constraint imposed by the Planck measurement of Thomson scattering would not be violated (Inayoshi et. al. 2016). Anyway, there may be scenarios in which some heating is produced while the reported large absorption is still possible.

\section{Population III stars, BH-HMXB-MQs, CRB, and BBHs}

UV radiation, X-rays, gamma-rays, gamma-ray bursts, and synchrotron jets can be produced by a plethora of sources: Pop III stars (e.g. Loeb 2010), gamma-ray bursts (e.g. Consumano et al. 2006), SN explosions (e.g. Furlanetto et al. 2004; Garrat-Smithson et al. 2018), Intermediate-Mass BHs (e.g. Madau et al. 2004; Ewall-Wice et al. 2018), Supermassive BHs (e.g. Biermann et al. 2014) and HMXBs (e.g. Mirabel 2011; Mirabel et al. 2011; Fragos et al. 2013). Some, if not all those sources may play some role at cosmic dawn, but a discussion of their relative roles is beyond the scope of this review. 

In the local universe it is observed that BH-HMXB-MQs are sources of hard X-rays and powerful synchrotron relativistic jets, which are frequently found in low metallicity dwarf galaxies. Therefore, it is natural to expect that BH-HMXB-MQs of Pop III are likely sources of a smooth synchrotron CRB. In fact, the most recent results from theoretical simulations show that  BH-HMXB-MQs are the immediate compact remnants of Pop III stars (Figure 5). 

\begin{figure}[h]
\begin{center}
 \includegraphics[width=5.3in]{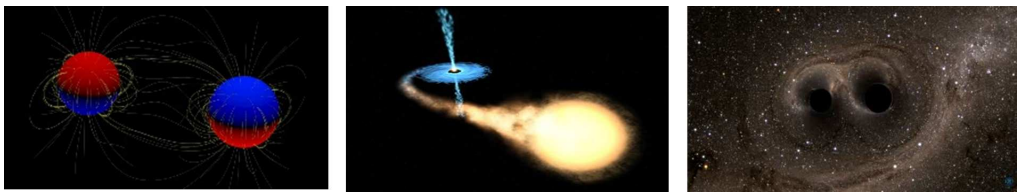} 
 \caption{ 
Binary Black Holes may be formed from relatively isolated massive stellar
binaries (Belczynski et al. 2016) or contact massive stellar binaries (De Mink \& Mandel 2016; Marchant et al. 2016), 
through an intermediate phase of BH-HMXB-MQs. Stars of $\geq$18-25 M$_{\odot}$ and metallicity Z$\leq$0.001 collapse directly or by failed SNe, and form BBHs that merge during Hubble time producing gravitational waves.}

   \label{fig5}
\end{center}
\end{figure}     

{\bf Massive stars of Pop III}. Early simulations had suggested that stars of Pop III have very large masses. However, with more powerful computational ability to follow the evolution of gas physics for adequately long periods of time at sufficiently high resolution, it was later found that the primordial gas is able to fragment. This fragmentation leads to a broad top-heavy Population III, extended to low masses ($\sim$1 M$_{\odot}$), but with the majority of stellar mass contained within the most massive stars of tens of solar masses, structured in stellar binaries and systems of larger multiplicity (Stacy et al. 2016 and references therein). 

Observations of stars in nearby Galactic open clusters by Sana et al. (2012) have shown that more than 70\% of all massive stars are in binaries or larger multiple systems. The fraction of systems with orbital periods $\leq$100 days is   
$\sim$85\% and the mass ratios of the binary components are in the range of 0.2 to 1. From those observations it was  concluded that the most common end product of massive-star formation is an interacting rather close binary, where 20 to 30\% of all O stars will merge. 

From a following more extensive observational campaign Sana et al. (2014) concluded that massive stars are formed nearly exclusively in multiple systems, and if corrected for observational biases, the true multiplicity fraction might very well come close to 100$\%$, despite different environments, sample ages and even metallicities. From a surprising dearth of short-period massive binaries in the very young ($\leq$1 Myr) massive star forming region M17, Sana et al. (2017) tentatively suggest that massive stars may be born with large orbital separations, and then harden or migrate to produce the typical (untruncated) power-law period distribution observed in few Myr-old OB binaries. 

{\bf BH-HMXB-MQs} are the next evolutionary state of primordial binaries with Z$\leq$0.001 and primary stars of $\geq$18-25 M$_{\odot}$. Stars of those low metallicities and high masses collapse directly or through failed SNe to form BHs (e.g. Heger et al. 2003), leaving the binary gravitationally bound as a BH-HMXB-MQ (Figure 5). 

BH-HMXB-MQs at cosmic dawn are the first steady high energy sources formed from the most massive stars of Pop III, and would  naturally produce a smooth synchrotron CRB. BH-HMXB-MQs of Pop III precede SN explosions, the formation of neutron stars, and the production of dust. They promptly inject hard X-rays and relativistic jets that overtake the slowly expanding HII regions, pre-heating the IGM over larger distance scales. 

A complete model of BH-HMXB-MQs formed from Pop III stars is being aimed by Sotomayor \& Romero (2018), in the context of a spectral energy distribution of the radiation produced by the accretion disk and the relativistic particles in the jets, in the framework of a lepto-hadronic model. The Pop III binary system
undergoes a BH-HMXB-MQ phase for $\sim$2x10$^{5}$years. During this lifetime, the donor star
loses $\sim$15 M$_{\odot}$, the BH mass remains
approximately constant, so almost all the accreted matter is 
ejected from the system in the form of jets \& massive outflows.

The BH accretes matter in a super-Eddington regime. This regimen of accretion
first theoretically examined by Shakura \& Sunyaev (1973) in a supercritical
accretion slim disk, was proposed for SS 433 by Fabrika (2004), and recently confirmed with NuSTAR by Middleton et al. (2018). Such steady super-Eddington accretion in SS433 may be a unique property among microquasars so far observed in the Milky Way.

{\bf BBHs} with components of 20-30 M$_{\odot}$ as the sources of gravitational waves detected by LIGO-Virgo (e.g. GW150914 \cite[Abbott et al. 2016]{Abbott_etal16}), can be formed by three main evolutionary channels as reviewed by Mirabel (2017): (1) BBHs formed from BH-HMXB-MQs by isolated evolution of massive stellar binaries (Belczynski et al., 2016), (2) BBHs formed from
BH-HMXB-MQs by the evolution of tight binaries with fully mixed chemistry (de Mink and Mandel, 2016;
Marchant et al., 2016), and (3) BBHs formed
by dynamical interaction in dense stellar clusters (Rodriguez et al. 2016 and references therein). 

In model (1) the BH members of BBHs are formed by direct collapse with no BH natal kicks that would unbind the binary. Model
(2) avoids the physics uncertainties in mass transfer, common
envelope mass ejection events, and the still unconstrained BH
kicks. However, model (2) assumes massive tight binary progenitors
of BBHs and has preference for BBHs with large masses
as in GW150914, but the relative low BH masses of $\leq$10  M$_{\odot}$ as seems to be the case in GW151226,
are difficult to reproduce in this model. In model (3) it is tacitly assumed that the members of BBHs
are also formed with no BH natal triggers that would eject
the BHs from the stellar cluster before BBH formation. 

BBHs formed by either
direct collapse and sufficiently low natal kicks from relatively isolated massive stellar binaries
(channel 1) or contact massive binaries (channel 2), will remain in
situ and ultimately merge in galactic disks. BBHs formed by dynamical
interactions in the cusps of dense stellar clusters, may be
ejected from their birth place and ultimately merge in galactic
haloes, like the sources of short gamma-ray bursts.

{\bf A gravitational wave background (GWB) from Population III BBHs} may be detectable by the future O5 LIGO/Virgo (Inayoshi et al. 2016). These authors argue that Pop III stars may form more BBHs but inject less ionizing photons into the IGM due to strong absorption by very high density of hydrogen. In that case, the GWB would be consistent with the cosmic early reionization by Pop III stars and the recent Planck measurement of Thomson scattering of 0.055$\pm$0.09.  Inayoshi et al. (2016) conclude that the detection of a flattening of the
spectral index of the GWB at frequencies as low as 30 Hz would be
an unique smoking gun of a high-chirp mass and high-redshift BBH
population expected from Pop III stars. 

Inayoshi et al (2017) studied the formation pathway of Pop III coalescing BBHs 
through stable mass transfer in Pop III binary stars without common envelope phases. For Pop III binaries with large and
small separations they find that $\sim$10\% of the total Pop III
binaries form BBHs only through stable mass transfer, and $\sim$10\% of these BBHs merge within the Hubble time. They conclude that the Pop III BBH formation scenario can explain
the mass-weighted merger rate of the LIGO’s O1 events with the maximal Pop III formation
efficiency inferred from the Planck measurement, even without BBHs formed by unstable
mass transfer or common envelope phases.

\section{Summary}

1) The theoretical and observational grounds for the hypothesis of a high formation rate of BH-HMXBs in the early universe (Mirabel 2011; Mirabel et al. 2011; Fragos et al. 2013) has been re-enforced in the last decade (e.g. Lehmer et al. 2016; section 1). 

2) Besides sources of hard X-rays and gamma-rays, HMXBs are also sources of powerful synchrotron relativistic jets and massive outflows, as revealed by observations of the best studied HMXBs in the Galaxy: Cygnus X-1, SS433 and Cygnus X-3 (section 2). 

3) Stellar black holes of $\geq$10 M$_{\odot}$ in the Milky Way (e.g. Cygnus X-1 \& GRS 1915+105), are formed by direct or failed SN collapse of massive stars of Z$\geq$Z${\odot}$. Therefore, it is expected that a significant fraction of Pop III stars of $\geq$18-25 M$_{\odot}$ and Z$\leq$0.001 Z${\odot}$ in binaries or larger multiple systems, collapse with no energetic SN kicks, remain in situ, ending as BH-HMXB-MQs of Pop III (section 3).   

4) BH-HMXB-MQs of Pop III are the most likely sources of a smooth synchrotron cosmic radio background (CRB). This CRB can provide an excess background temperature (Feng \& Taylor 2018) to boost the deep absorption of HI centered at z$\sim$17 reported by EDGES (Bowman et al. 2018a). 

5) Theoretical models of the tomography of atomic hydrogen at cosmic dawn that  incorporate heating of the IGM by sources  with a hard X-ray spectral distribution like BH-HMXBs  (e.g. Furlanetto 2006; Mirabel et al. 2011, Cohen et al. 2017), predict the mean redshift (z$\sim$17) of the deep and frequency wide (z$\sim$14 to 21) HI absorption reported by EDGES, for UV and X-ray escape values $\leq$0.1. This implies column depths N$_{H}$$\geq$5$\times$10$^{23}$ cm$^{-2}$ (Ewall-Wice et al. 2018) for the cold hydrogen IGM that enshrouds the high energy sources. These column depths block the X-rays but are transparent for the synchrotron CRB (section 6).  

6) Massive stars in the Milky Way are formed nearly exclusively in multiple systems, despite different environments, sample ages and even metallicities (Sana et al. 2014). From theoretical simulations it is concluded that population III stars are of broad top-heavy masses, extended to low masses ($\sim$1 M$_{\odot}$), with the majority of stellar mass contained within the most massive stars of tens of solar masses, structured in stellar binaries and systems of larger multiplicity (Stacy et al. 2016; section 7).  

7) BH-HMXB-MQs in stellar clusters of Pop III are formed before the appearance of SN explosions, neutron stars and dust, which would be consistent with an absence of a cosmic far-infrared thermal background associated with a Cosmic Radio Background. BH-HMXB-MQs promptly inject hard X-rays and relativistic jets in the cold hydrogen that enshrouds the slowly expanding HII regions ionized by the most massive progenitor stars of Pop III (section 7). 

8) BH-HMXB-MQs of Pop III would be different to most known BH-MQs in the Milky Way. The steady super-Eddington accreting source SS433 may be the only observed Galactic analog of BH-HMXB-MQs of Pop III (Sotomayor \& Romero 2019; Section 7).

9) The detection of a flattening of the
spectral index of the Gravitational Wave Background at frequencies as low as $\sim$30 Hz (which corresponds to merging BHs of 30-40 M$_{\odot}$), would be
the smoking gun of a high-chirp mass, high-redshift BH-HMXBs and BBHs
populations, expected from Pop III stars (Inayoshi et al. 2016; section 7)

10) Future observations with more sensitive X-ray missions (e.g. Athena) will allow more precise quantitative determinations to test the impact of HMXBs in Cosmology. The future enhanced statistics of sources of Gravitational Waves (GWs) by GW observatories (e.g. LIGO-Virgo) will provide larger data bases to test the cosmic evolution of HMXBs and BBHs.

{\bf Acknowledgments:} I thank Dale Fixsen for comments on ARCADE 2 results, Bret Lehmer, Jayce Dowel, Gregory Taylor and Judd Bowman for permission to use their published figures.

\end{document}